# Training atomic neural networks using fragment-based data generated in virtual reality

Silvia Amabilino,[1,2] Lars A. Bratholm,[1,2] Simon J. Bennie,[1,2] Michael B. O'Connor,[2,3] David R. Glowacki[1,2,3*]

[1]*School of Chemistry, University of Bristol, Bristol, BS8 1TS, UK;* [2]*Intangible Realities Laboratory, University of Bristol, BS8 1UB, UK;* [3]*Department of Computer Science, University of Bristol, BS8 1UB, UK*
[*]*glowacki@bristol.ac.uk*

## Abstract

The ability to understand and engineer molecular structures relies on having accurate descriptions of the energy as a function of atomic coordinates. Here we outline a new paradigm for deriving energy functions of hyperdimensional molecular systems, which involves generating data for low-dimensional systems in virtual reality (VR) to then efficiently train atomic neural networks (ANNs). This generates high-quality data for specific areas of interest within the hyperdimensional space that characterizes a molecule's potential energy surface (PES). We demonstrate the utility of this approach by gathering data within VR to train ANNs on chemical reactions involving fewer than 8 heavy atoms. This strategy enables us to predict the energies of much higher-dimensional systems, e.g. containing nearly 100 atoms. Training on datasets containing only 15k geometries, this approach generates mean absolute errors around 2 kcal mol$^{-1}$. This represents one of the first times that an ANN-PES for a large reactive radical has been generated using such a small dataset. Our results suggest VR enables the intelligent curation of high-quality data, which accelerates the learning process.

# 1. Introduction

In the recent past, computations were mostly limited by the available processing power. With the machine learning revolution, the issues related to generating and curating data have become equally as important as the algorithms used to process and learn the data[1].

The molecular sciences have seen a surge in popularity of machine learning methods for a variety of applications, from designing new drug molecules[2, 3] to planning synthetic chemistry strategies[4]. Multiple research groups have been applying machine learning to the prediction of molecular energies and forces[5-8], with the goal of accelerating molecular dynamics (MD) simulations. For small systems, ab initio calculations can be used to evaluate accurate energies and forces at each step of an MD simulation. However, this becomes too computationally expensive for larger systems and more approximate methods, such as force fields, are generally used. While much faster, these incur a trade-off in accuracy.

An alternative to force fields is to use accurate data, e.g. from electronic structure calculations, to fit potential energy surfaces (PES). Evaluating fitted PES should be faster than performing electronic structure calculations but should provide similar accuracy to the underlying data. A variety of machine learning techniques have been used for either fitting or interpolating PES, for example permutationally invariant fitting[9], cubic splines[10], modified Shepard interpolation[11], interpolating moving least squares[12], Multi-State Empirical Valence Bond theory (MS-EVB)[13, 14], reproducing Kernel Hilbert space interpolation[15]. Recently, Kernel Ridge Regression (KRR) and Neural Networks (NNs) have attracted considerable attention[16, 17]. However, while fast to train, KRR suffers from poor scaling with data set size, while NNs memory scaling does not depend on the number of data points. This contributed to NNs popularity for applications with large datasets.

In 2007, Behler and Parinello introduced atomic neural networks and Atom Centred Symmetry Functions (ACSFs).[7] They used them to fit the potential energy surface of bulk silicon, and showed how simulations using the fitted potential could accurately reproduce the DFT radial distribution of melted silicon at 3000 K. In 2012, Artrith and Behler used atomic neural networks with ACSFs to study copper surfaces.[18] Their method was implemented in a closed source software called RuNNer.[19] The combination of atomic neural networks with ACSFs started becoming more popular after the paper by Behler titled 'Constructing High-Dimensional Neural Network Potentials: A Tutorial Review'.[20] In the years following this publication, multiple groups started working on their own implementation of these neural network models[1, 5, 6, 21].

In 2017, NNs Smith et al. used NNs to fit quantum mechanical DFT calculations in order to learn accurate and transferable potentials for organic molecules with up to 8 heavy atoms[5]. These NNs were able to predict the energies of larger molecules with up to 53 atoms. However, they were trained on a dataset containing around 17.2 million compounds. Similarly, Kun et al. showed that by training on 15k different molecules and 3 million geometries, they obtained low errors when predicting the energy of molecules outside of the dataset[6]. Generating these large datasets can be extremely computationally expensive if accurate energies are required. Furthermore, sampling data to study the dynamics of reactive systems is especially challenging. A previous study[1] investigated how real-time interactive quantum molecular dynamics in virtual reality (iMD-VR) can be used to generate data for training NNs. iMD-VR relies on human intuition to efficiently sample hyperdimensional PESs[22], as users can interact 'on-the-fly' with real-time quantum mechanical molecular dynamics simulations and explore regions of interest on the PES[1]. The NN trained on iMD-VR data was found to have similar performance to one trained on constrained molecular dynamics data, but the former had a lower mean absolute error (MAE) when predicting the energies of geometries close to the minimum energy path of the reaction.[1]

In this article, we investigate how real-time iMD-VR can be combined with fragment-based training of NNs to predict the energy of large open-shell reactive systems accurately. We focus on the reaction of a large hydrocarbon chain ($C_{30}H_{62}$, called 'squalane') and a cyano radical (CN). We use a training dataset that does not contain the system we want to predict, but only smaller hydrocarbons with up to 6 carbon atoms.
To date, most studies of chemical reaction surfaces with NNs have focused on systems with up to 19 atoms[1]. Therefore, with 94 atoms, this system represents (to the best of our knowledge) one of the largest radical systems for which an NN-fitted PES has been developed.

# 2. Methods

Creating a data set for a system as large as squalane can be extremely time-consuming. First of all, sampling a large number of geometries is harder than for smaller systems, such as smaller hydrocarbons. If the sampling is based on molecular dynamics, evaluating the forces acting on each atom at each time step of the simulation takes longer due to the larger number of atoms. In addition, refining the energies of each geometry with a more accurate electronic structure calculation takes longer. Consequently, instead of training the NN on various conformations of the system of interest (i.e. squalane reacting with the cyano radical), we decided to train it on a data set consisting of smaller hydrocarbons (i.e. methane, ethane, isobutane, isopentane and isohexane) reacting with cyano radical. This enabled us to investigate further the transferability of atomic NNs.

Atomic NNs have been shown in the literature to be transferable when learning the potential energy surface of small organic molecules around geometries corresponding to energy minima[5]. Here we build on these findings by studying if atomic NN potentials are also transferable for reactive systems. Transferability for reactive systems allow the study of reaction mechanisms at a high level of theory, without having to generate high-quality data for the large systems. Since expensive calculations would only be needed for smaller fragments of the system, this may enable to study reactive systems that have so far been out of reach due to computational constraints.

## 2a. Generating the data sets

The data sets were generated using iMD-VR within the Narupa framework[23] for hydrocarbons of a variety of lengths. One hydrocarbon and the cyano radical were then loaded in the Narupa environment and spawned in random non-overlapping positions. The hydrocarbons used were methane, ethane, isobutane, isopentane, 2-isohexane, 3-isohexane and squalane. For each hydrocarbon, the sampling of the reaction with the cyano radical was performed by first loading a starting structure of each of the reactants in XYZ format into the Narupa environment. They were spawned in random (non-overlapping) positions within a cubic box with length 30 Å. In order to sample hydrogen abstractions, the reactants were brought in proximity to each other by the user, to enable the reaction to take place (Figure 1 shows this process for isopentane).

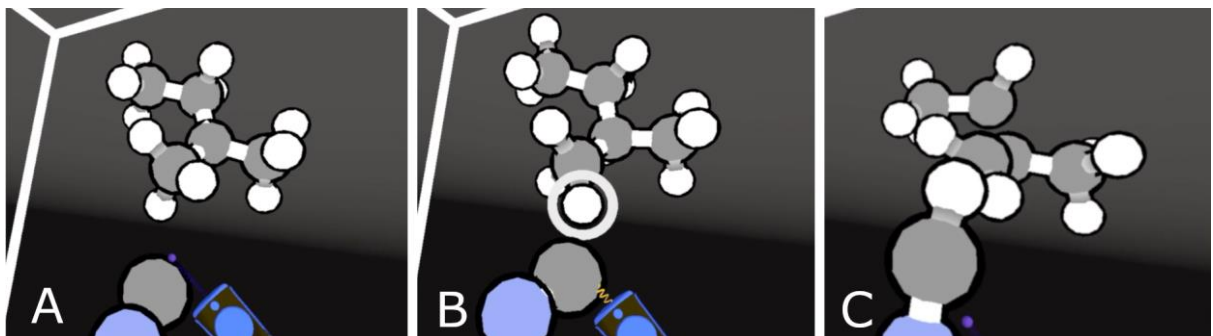

*Figure 1 - Reaction of isopentane and cyano radical sampled in iMD-VR. The reactants (A) are brought in close proximity (B) to form the products (C).*

Once the products were formed, they were pulled away from each other. After each reaction, the system was re-initialized to a random configuration of the reactants. It was found that separating each trajectory in this way made the data processing considerably easier, because it is possible to keep track from which trajectory each configuration came from.

The molecular dynamics simulations were run using a Velocity Verlet integrator with a time step of 0.5 fs. The Andersen thermostat was used to maintain the system temperature at 300 K, with a collision frequency of 10 $p^{-1}$ $s^{-1}$. The system was constrained to stay within the box via velocity inversion. For the interaction of the user with the atoms, a spring potential with a force constant of 1000 kJ $mol^{-1}$ $Da^{-1}$ was used. A velocity re-initialization procedure was used to rapidly re-equilibrate the system between interactions with the user. We used the semi-empirical method PM6[24] to evaluate energies and forces in the molecular dynamics simulation. The implementation of PM6 is from the SCINE sparrow package (http://scine.ethz.ch), developed by Reiher and co-workers[25-28]. This package includes implementations of tight-binding engines like DFTB alongside a suite of other semi-empirical methods.

For methane reacting with CN, 81439 configurations were sampled in iMD-VR and were then reduced to 18000. This was done by taking the average of the first 400 frames and the last 400 frames to find the energies of reactants and products. Then, at most 600 frames before and after the mid-point between the reactants and products energies were taken. For ethane, the same procedure was followed. In this case, 28969 configurations were obtained with iMD-VR and were then reduced to 7975. For isobutane, 40984 samples were sampled and then pruned to 13113. For isopentane, we reused the iMD-VR data set obtained for our previous publication[1]. Two different isomers of hexane were sampled: 2-isohexane and 3-isohexane. For 2-isohexane, 36389 configurations were initially sampled and they were then pruned using the same procedure as for methane, but in addition the structures with energy 113 kJ $mol^{-1}$ higher than the energy of the reactants were removed. This left 13084 samples, with a total of 11 abstractions. For 3-isohexane, 45179 samples were obtained and then pruned similarly to 2-isohexane. This left 13 trajectories for a total of 14912 samples.
The final molecular system sampled was squalane reacting with the cyano radical. Only one trajectory was generated, as the SCINE implementation used in iMD-VR had not yet been parallelized, and therefore resulted in about 1 time step per second being rendered in iMD-VR. This made it extremely difficult to effectively bias the sampling. It should be noted that since then, the implementation of PM6 in Narupa has been improved and is parallelized. Consequently, it is now easier to sample large systems.

In summary, the final data set contained 15 trajectories for methane, 8 trajectories for ethane, 11 trajectories for isobutane (7 primary and 4 tertiary), 19 trajectories for isopentane (11 primary, 3 secondary and 5 tertiary abstractions), 11 trajectories for 2-isohexane (5 primary, 4 secondary and 2 tertiary), 13 trajectories for 3-isohexane (6 primary, 3 secondary and 4 tertiary) and one trajectory for squalane (secondary).

The energies of all configurations were computed with electronic structure calculations using MOLPRO[29]. The Coulomb fitted[30] unrestricted PBE[31] functional with the Def2-TZVP[32] basis set were used. In the rest of the report this method is referred to as CF-uPBE/TZVP. We used this level of theory to remain consistent with our previous study[1]. The DFT energies of the geometries from the squalane abstraction trajectory are shown in Figure 2.

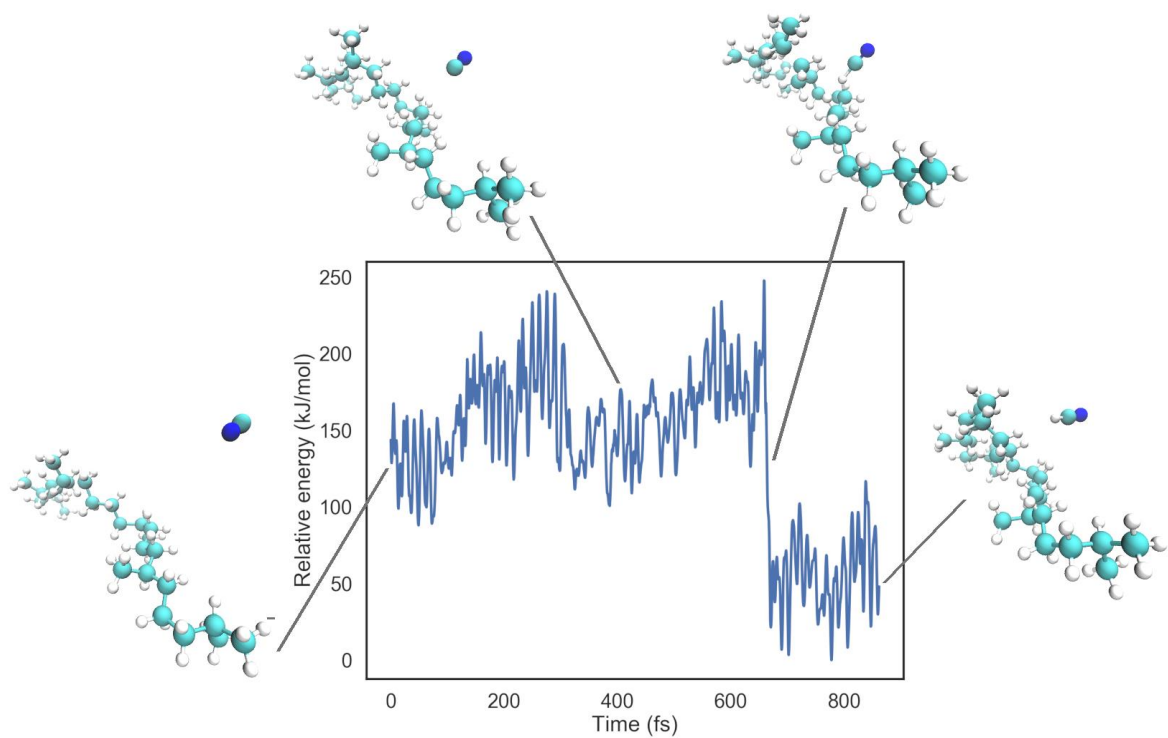

*Figure 2 - Cyano radical abstracting a secondary hydrogen from squalane. It shows how the energy evolves over time. The energies are shown relative to the geometry with lowest energy in the trajectory.*

The energies of the different hydrocarbons have considerably different magnitudes. We found that the NN training converges faster if the data is pre-processed by shifting the energy values to be in a smaller range. We shifted the total energy of each configuration by subtracting a constant value for each atom dependent on its type. The contribution of each element to the total energy was obtained with a Lasso linear model[33]:

$$\mathbf{y} = \mathbf{X}\boldsymbol{\beta} + \boldsymbol{\varepsilon}$$

where $\mathbf{y}$ is the vector of the energies of the different hydrocarbons and cyano radical, $\mathbf{X}$ is a matrix of the features for all the samples (the number of C, H and N atoms in each sample), and $\boldsymbol{\varepsilon}$ is a constant and $\boldsymbol{\beta}$ is the vector of regression coefficients. The regression coefficient is regularised with L1 regularisation during the optimisation.

To fit the Lasso model, 100 configurations of the reactants of each system were used. The number 100 was chosen because it seemed enough to represent the average energy of the reactants. The $\mathbf{X}$ matrix has dimensions $(N, M)$, where $N$ is the number of samples and $M$ is the number of element types present in the systems. In this case, $M = 3$ since there are only

H, C and N atoms present in all configurations. This means that a row of **X** corresponding to a configuration of $CH_4$ + CN is $\mathbf{X_i}$ = [4, 2, 1], as there are 4 H atoms, 2 C atoms and 1 N atom. $\mathbf{y_i}$ is the energy of the $\mathbf{X_i}$ sample. Once the Lasso is fit, it can predict a shifting factor for hydrocarbon systems with different numbers of atoms. This shifting factor was subtracted from the energies in the training set. The shifting of the energies was done in this way instead of using the atomisation energies so that some of the bonding energy can also be accounted for.

Six mixed data sets with the same total number of data points (15000) but different ratios of the various species were constructed. The composition of the six data sets is shown in Table 1. For each mixed data set, at least one full trajectory was left out the training for each different hydrocarbon, so that it could be used for testing.

*Table 1 - Composition of the six different training sets used to train the atomic NNs.*

| Training set | Composition |
| --- | --- |
| 1 | 15000 Methane |
| 2 | 10000 Methane |
|   | 5000 Ethane |
| 3 | 8000 Methane |
|   | 4000 Ethane |
|   | 3000 Isobutane |
| 4 | 7500 Methane |
|   | 3500 Ethane |
|   | 2500 Isobutane |
|   | 1500 Isopentane |
| 5 | 7500 Methane |
|   | 7500 Isopentane |
| 6 | 8000 Methane |
|   | 4000 Isopentane |
|   | 1500 2-Isohexane |
|   | 1500 3-Isohexane |

## 2b. Training the NN

Similarly to our previous work[1], we used the atomic NNs architecture [34] and the Smith formulation of the Atom Centred Symmetry Functions[5] (ACSFs) as the atomic representation [35]. We used the implementation of atomic NNs and ACSFs present in the QML Python package[35].

After generating the datasets, we optimised the hyper-parameters of both the ACSFs and the for each dataset using a random search[36] using group 3-fold cross validation[1]. The values of the optimal hyper-parameters obtained are shown in the Supporting Information.

After training, the performance of each atomic NN was evaluated by predicting the energy of squalane ($C_{30}H_{62}$) reacting with CN and comparing it to the reference DFT data. A constant offset was removed from all atomic NNs predictions in order to get a better comparison of the quality of the relative predictions.

The scripts used to train the atomic NNs and to generate the figures can be found on Github (https://github.com/SilviaAmAm/squalane_paper_si).

# 3. Results and discussion

## 3a. The data sets

Figure 3 shows the $C_{hydrocarbon}$-H and $C_{CN}$-H distances (for the hydrogen being abstracted) during the sampled abstraction trajectories of isobutane and squalane (after pruning). As can be seen from Figure 3, the structures sampled the most have either a short $C_{hydrocarbon}$-H or a short $C_{CN}$-H distance. These are reactant and product structures. Transition state structures, where the distance between the hydrocarbon C and the H is greater than 1.2 Å and the cyano C and the H is greater than 1.1 Å, are not sampled as often. This is evident from the lighter colour of the plot, due to the lower density of points in this region. Obtaining additional samples near the transition state configuration would require sampling many more trajectories in iMD-VR. Then, most of the data points near equilibrium would be discarded, while most of the data points near the transition state would be kept. This would give a more homogeneous distribution of samples. However, it would be considerably more time-consuming.

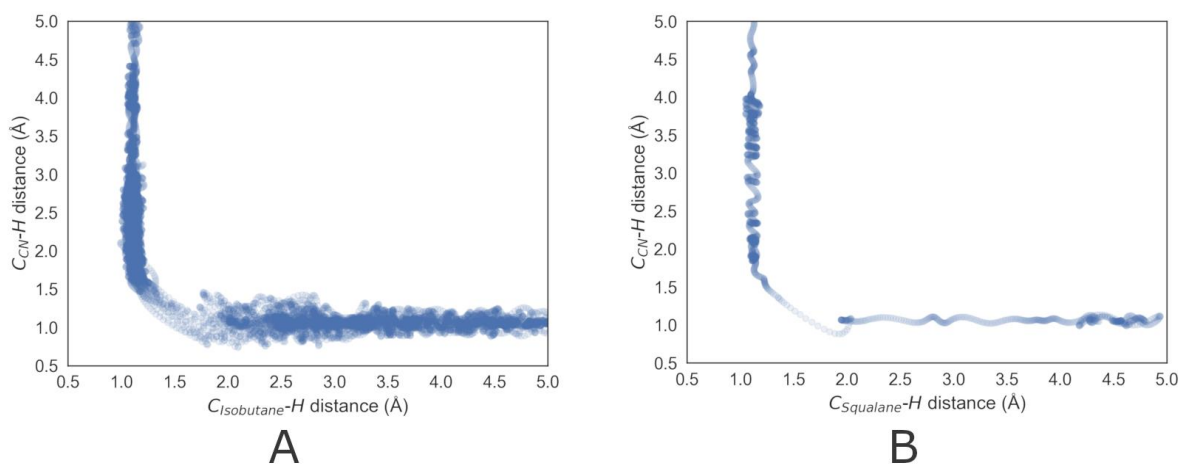

*Figure 3 - Distances between the cyano carbon and the abstracted hydrogen as a function of the distance between the hydrocarbons carbon and the abstracted hydrogen. This is shown for A) isobutane and B) squalane. Each data point is plotted with transparency, so that the difference in the sampling of various regions can be observed.*

After computing the energies at the CF-uPBE/TZVP level, they were shifted using the Lasso model. The energies before and after shifting are shown in Figure 4. Only 5000 data points per hydrocarbon are shown. The average energy of the methane and cyano radical system was used as the reference energy. The longer the hydrocarbon, the lower the energy of the system. Figure 4A shows a 'zoomed out' view of all the energies. Since the range is of around $3 \times 10^6$ kJ mol$^{-1}$, the details of the trajectories cannot be seen. Figure 4B shows a 'zoomed in' view of the trajectories, where the energy range is only around $2 \times 10^2$ kJ mol$^{-1}$. This shows how for larger hydrocarbons, the difference in energy between the reactants and the products is larger.

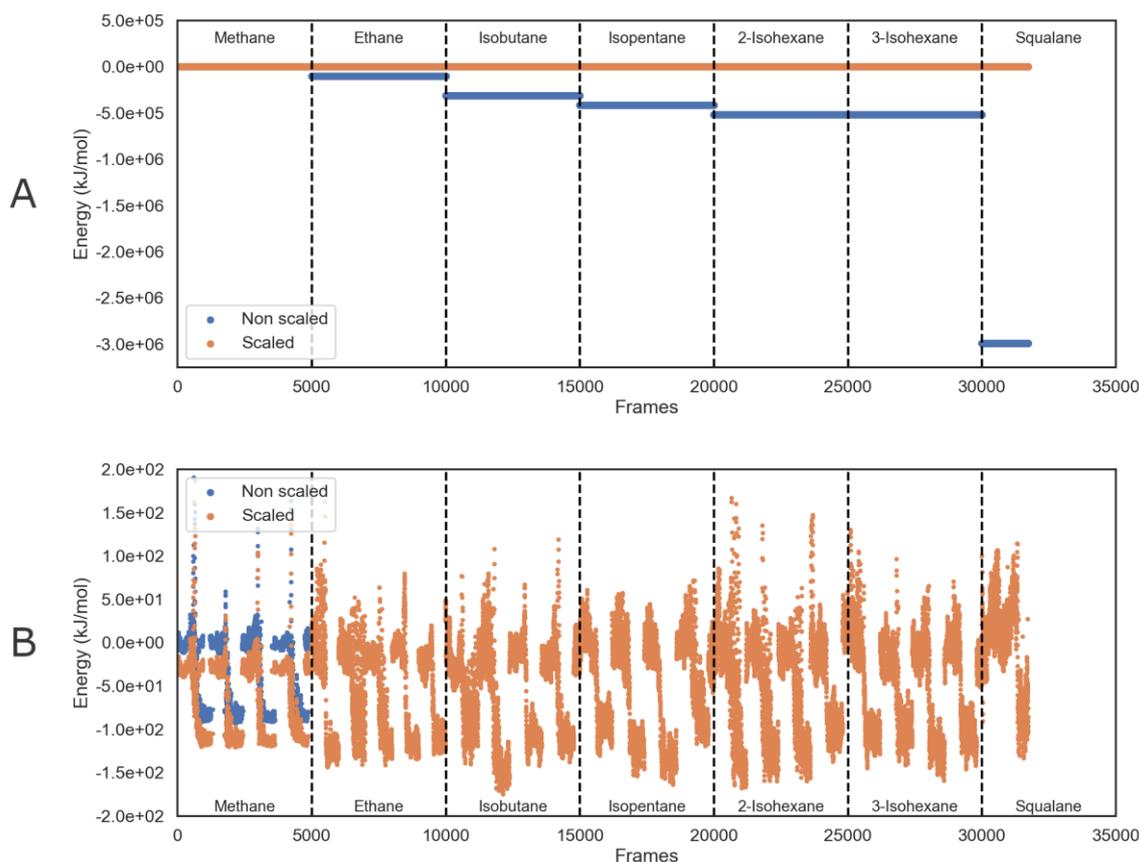

*Figure 4 - Energies of all the systems with different sizes of hydrocarbons before (blue) and after (orange) shifting with the Lasso model. Part B is a zoomed in view of part A in the range of the scaled energies.*

### 3b. Fitting the squalane + CN surface

An atomic NN was trained for each mixed data set with the highest scoring set of hyper-parameters obtained from the hyper-parameter optimisation. In order to compare the relative performance of the NNs trained on the six different data sets, the mean absolute errors (MAEs) and the coefficient of determination ($R^2$) of the predictions were calculated (Table 2). $R^2$ was calculated with the scikit-learn $R^2$ score function, which can give negative numbers[37].

Table 2 - *Mean Absolute Errors and $R^2$ values of the atomic NNs predictions for the trajectory of squalane reacting with CN compared to the DFT reference.*

| Trained on | MAE (kJ mol$^{-1}$) | $R^2$ |
|---|---|---|
| Training set 1 | 41.52 ± 30.53 | 0.01 |
| Training set 2 | 43.20 ± 35.91 | -0.18 |
| Training set 3 | 20.68 ± 17.01 | 0.73 |
| Training set 4 | 9.87 ± 9.26 | 0.93 |
| Training set 5 | 11.54 ± 8.53 | 0.92 |
| Training set 6 | 8.16 ± 6.9 | 0.96 |

The first NN was trained on training set 1, which contained 15000 samples of methane and CN. It was then used to predict the energy of the squalane test trajectory (Figure 2). The MAEs of the predictions are reported in Table 2 and the predicted energies for the squalane abstraction trajectory are shown in Figure 5. While the NN trained on training set 1 can predict the energies of methane reacting with CN very accurately, it does not generalise to squalane.

Figure 5 shows that the predicted trajectory does not show a change in energy between the reactants and the products and the changes in energies due to the stretching motions of the bonds are not well reproduced. The poor performance of this NN in predicting squalane is not surprising, since it could not have learnt about carbon-carbon bonds and secondary hydrogen abstractions.

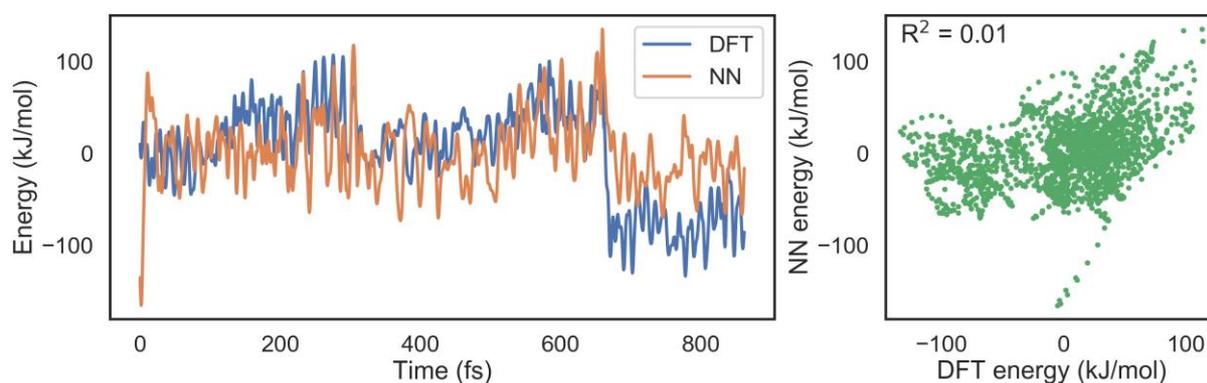

*Figure 5 - NN trained on training set 1 (containing only methane reacting with CN) predicting the energies of the trajectory of squalane reacting with CN.*

There is an improvement once ethane is added to the training set. The NN trained on training set 2 gave a trajectory with a clear difference between the predicted energies of the reactants and the products (Figure 6). However, the changes in energies due to the stretching motions of the bonds are still not captured.

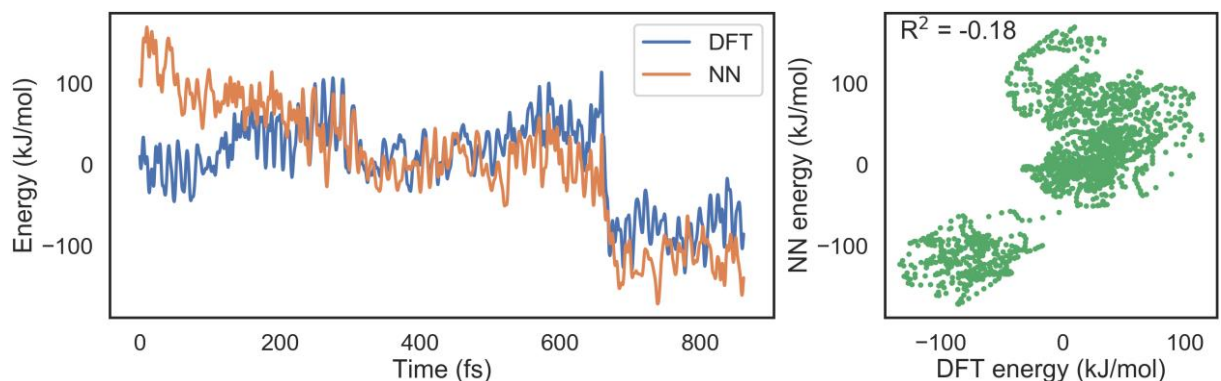

*Figure 6 - NN trained on training set 2 (containing methane and ethane reacting with CN) predicting the energies of the trajectory of squalane reacting with CN.*

When isobutane is added to the training set (training set 3), the NN can learn about tertiary hydrogens. The correlation plot for the energy predictions and the DFT energies starts to approach a straight line (green plot in Figure 7). It appears that the energies of the products are predicted worse compared to those of the reactants. This is probably because in isobutane only primary and tertiary abstractions can be sampled, but the squalane test trajectory is a secondary abstraction.

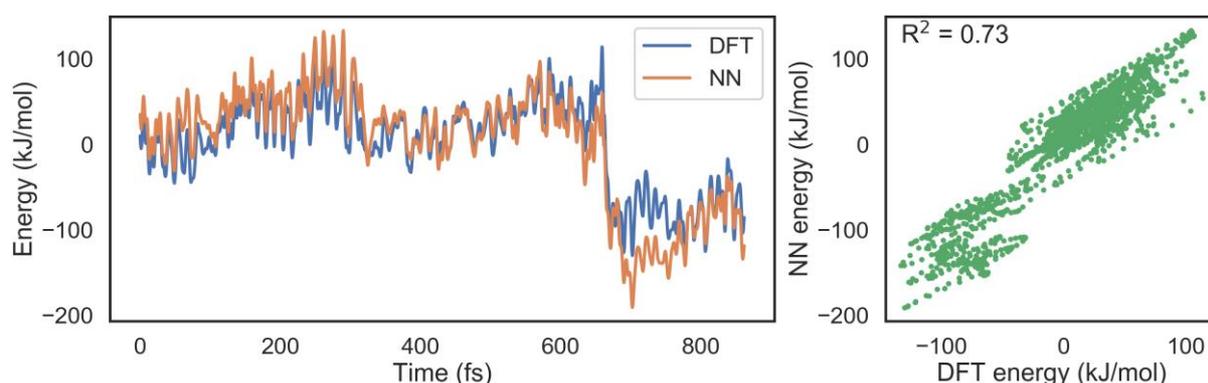

*Figure 7 - NN trained on training set 3 (containing methane, ethane and isobutane reacting with CN) predicting the energies of the trajectory of squalane reacting with CN.*

When the training set contains 7500 methane, 3500 ethane, 2500 isobutane and 1500 isopentane (training set 4), the model can finally learn about secondary hydrogen abstractions. Now the oscillations in energy due to the stretching motions of the bonds are better captured and the energy difference between the products and the reactants is similar to that seen in the reference data (Figure 8).

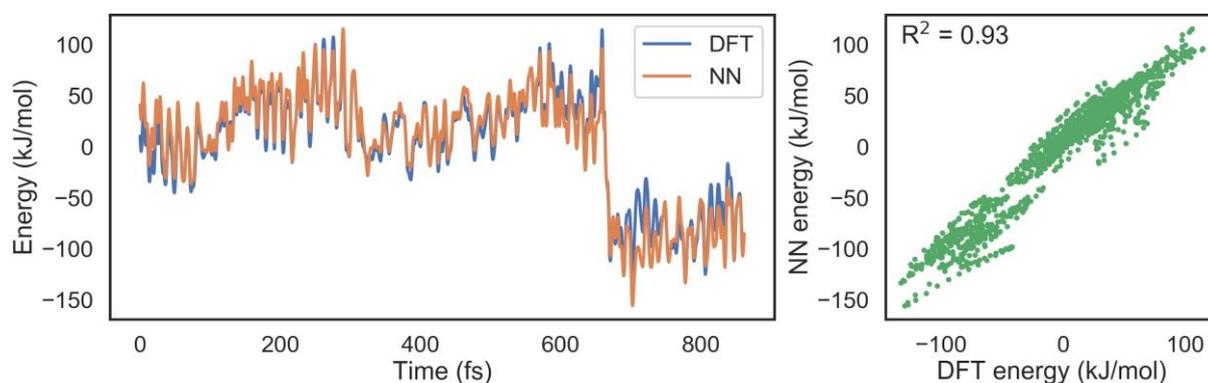

*Figure 8 - NN trained on training set 4 (containing methane, ethane, isobutane and isopentane reacting with CN) predicting the energies of the trajectory of squalane reacting with CN.*

The NN trained on the data set with 7500 methane and 7500 isopentane data points (training set 5) gave very similar results to the one with methane, ethane, isobutane and isopentane (training set 4), which suggests that the shorter hydrocarbons still contribute valuable information and reduce the number of isopentane samples required to obtain a good description of squalane (Figure 9).

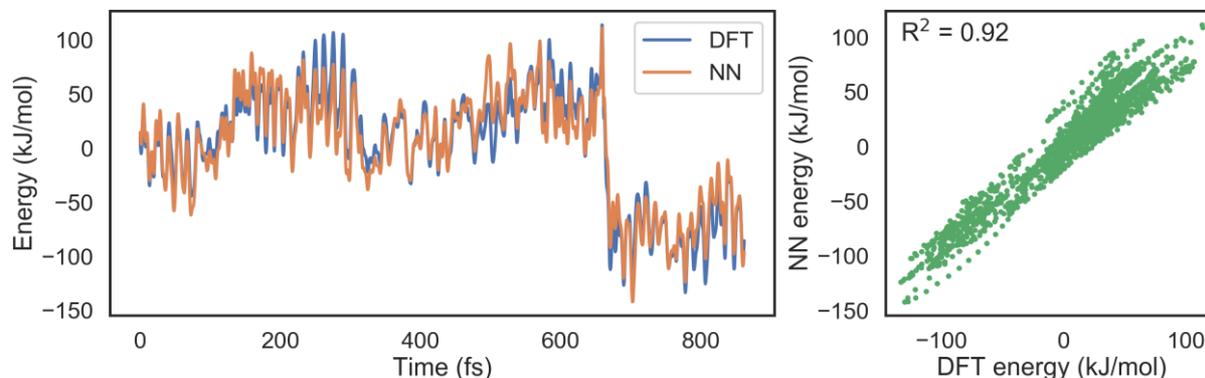

*Figure 9 - NN trained on training set 4 (containing methane and isopentane reacting with CN) predicting the energies of the trajectory of squalane reacting with CN.*

When larger hydrocarbons such as isohexane are added to the training set (training set 6), the performance of the NN improves only slightly (Figure 10). This suggests that isopentane contains enough information to learn the most important features of the potential energy surface and the addition of isohexane does not improve significantly to the results.

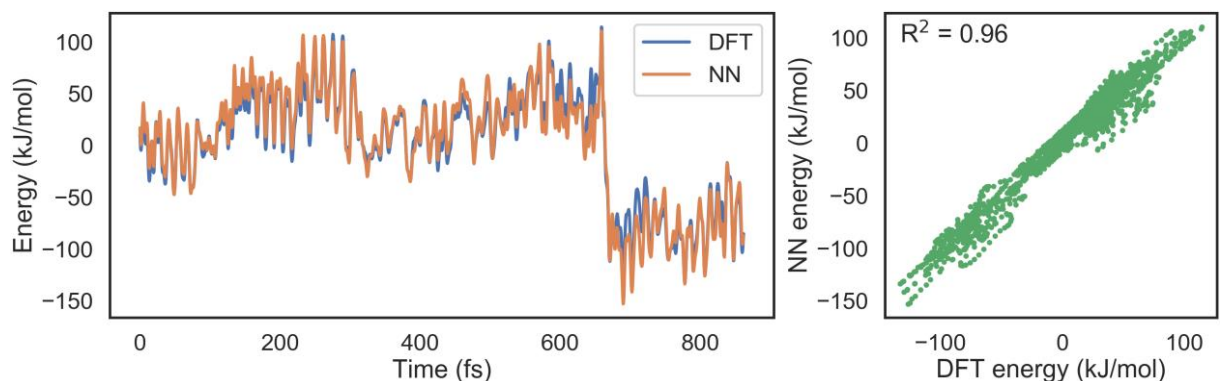

*Figure 10 - NN trained on training set 6 (containing methane, isopentane and isohexane reacting with CN) predicting the energies of the trajectory of squalane reacting with CN.*

### 3c. Comparison with PM6 energies

The predictions of the NN trained on the data set containing isopentane and isohexane (training set 6) were compared to the PM6 energies for the squalane trajectory (Figure 11). A constant was removed from the PM6 energies in the same way as for the predictions for the NNs, in order to minimise the error from the DFT energies. The PM6 energies have an MAE of 34.8 kJ mol$^{-1}$ compared to the DFT energies, while the predictions of the NN trained on training set 6 only have an MAE of 8.2 kJ mol$^{-1}$. This is also evident from Figure 11, where the NN predictions are closer to the DFT energies compared to the PM6 energies, especially for the products. The energy of the squalane radical and HCN is under-estimated by PM6 by about 50 kJ mol$^{-1}$.

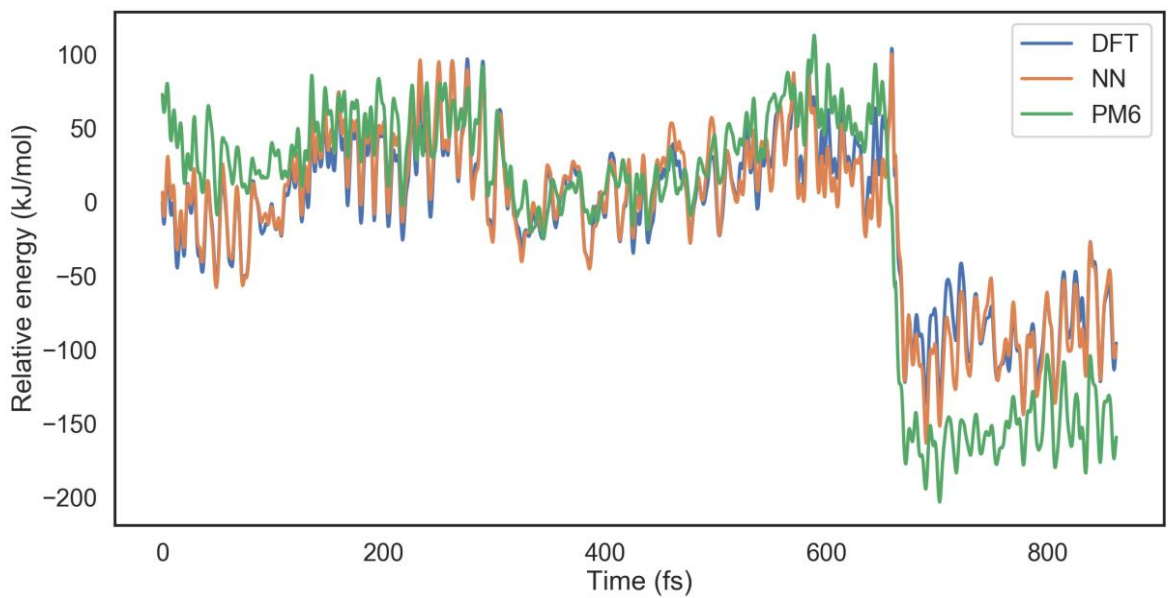

*Figure 11 - Comparison of the predictions from the NN trained on training set 6, the PM6 energies and the DFT energies of the trajectory of squalane reacting with CN. Energy offsets have been removed for both the NN predictions and the PM6 energies, in order to give a fair comparison.*

We also compared the speed of obtaining the energies from PM6 and the NNs. The time measurements depend heavily on the hardware used. Here we used a Nvidia GEFORCE GTX 1080Ti.

Since all the NNs presented here had different hyper-parameters, the speed of prediction for each one of them was different. The hyper-parameters that influence the speed of prediction the most are the number of features of the ACSFs and the number of hidden neurons in the hidden layers.
When using a NN to predict the energy of a configuration, there are two steps. The first is to evaluate the representation (in this case the ACSF). Once the representation has been generated, it is input into the NN which outputs the total energy. Since the time taken for each of these steps depends on the hyper-parameters, the measurements for the representation and for the energy prediction are shown separately in Table 3, as well as the combined time. The results show that the average timings are in the order of 10 ms for each sample.

*Table 3 - Average times obtained when evaluating the energy of 1725 CN + squalane configurations using NNs trained on the 6 mixed data sets. The timings are divided in the ACSFs generation and the energy prediction. The total time for each data sample is shown.*

| NN trained on training set | ACSF time (ms) | Energy time (ms) | Total time (ms) |
|---|---|---|---|
| 1 | 7 | 1 | 8 |
| 2 | 8 | 2 | 10 |
| 3 | 4 | 1 | 5 |
| 4 | 4 | 1 | 5 |
| 5 | 4 | 1 | 5 |
| 6 | 6 | 1 | 7 |

DFT (CF-uPBE/TZVP) was used to calculate the energies of the 1725 squalane configurations (MOLPRO was used, without parallelisation). This took on average in the order of 1000 s per configuration. On the other hand, PM6 took in the order of 0.1 s. The comparison to the NN timings is shown in Table 4. This is an encouraging results, since the NN appears to be the fastest at predicting the energies, while giving better energy results compared to PM6 (Figure 11).

*Table 4 - Comparison of the order of magnitudes of the average timings for evaluating the energy of a squalane + CN geometry using NNs, CF-uPBE/TZVP (DFT) and PM6.*

| Method | Time (s) |
| --- | --- |
| NN | 0.01 |
| DFT (CF-uPBE/TZVP) | 1000 |
| PM6 | 0.1 |

## 4. Conclusions

This study investigated the transferability of reactive potential energy surfaces (PES) fitted with atomic neural networks, with training data generated using interactive molecular dynamics in virtual reality.

The atomic NNs were trained on six different training sets containing different proportions of small hydrocarbons (with at most 6 carbon atoms) reacting with a cyano radical (CN). After training, the atomic NNs were used to predict the energy of squalane ($C_{30}H_{62}$) reacting with CN.

The results showed that only including methane and ethane in the training set gives poor results. Already when using training set 3 (containing methane, ethane and isobutane), the mean absolute errors are around 20 kJ mol$^{-1}$. The results improve considerably once isopentane is added to the training set (MAE around 10 kJ mol$^{-1}$), but then there is only around 2 kJ mol$^{-1}$ improvement if longer hydrocarbons are added.

In conclusion, this shows that reactive potential energy surfaces fitted with atomic NNs can be transferable. However, this is the case only if enough molecules in the training set capture the key chemical interactions of the large system. Furthermore, we showed that good accuracy can be achieved with training sets containing only 15000 data point. Following this approach, the cost of generating the dataset and training the neural network can be drastically reduced.

## Acknowledgements

D.R.G. acknowledges funding from the Royal Society as a Univ. Research Fellow and also from EPSRC Program Grant No. EP/P021123/1. L.A.B. acknowledges funding from EPSRC Program Grant No. EP/P021123/1. LAB thanks the Alan Turing Institute under the EPSRC grant EP/N510129/1. Funding for S.A. is from the EPSRC Centre for Doctoral training, Theory and Modelling in Chemical Sciences, under Grant No. EP/L015722/1. S.J.B. thanks the BBSRC and the Royal Society of Edinburgh Enterprise fund, EPSRC for Grant No. EP/M022129/1, and the UoB School of Chemistry for funding. We also acknowledge helpful conversations throughout with Prof. M. Costen, Prof. K. McKendrick, and Dr. S. Greaves from Heriot-Watt Univ. (Edinburgh) while preparing this draft.

We further acknowledge the following programs: Open Babel[38], Pybel[39], Numpy[40], Matplotlib[41], VMD[42], Avogadro[43], OpenMP[44], F2PY[45] and TensorFlow[46].

# Supporting Information for Publication

# Training atomic neural networks using fragment-based data generated in virtual reality

Silvia Amabilino,[1,2] Lars A. Bratholm,[1,2] Simon J. Bennie,[1,2] Michael B. O'Connor,[2,3] David R. Glowacki[1,2,3*]

[1]School of Chemistry, University of Bristol, Bristol, BS8 1TS, UK; [2]Intangible Realities Laboratory, University of Bristol, BS8 1UB, UK; [3]Department of Computer Science, University of Bristol, BS8 1UB, UK
*glowacki@bristol.ac.uk

## S1. Hyper-parameters used for training the ANNs

The hyper-parameters for the NNs trained on each mixed data set are shown in Table 1. The hyper-parameter shown in the table are the keywords used in the QML package. Their meaning is:

- `iterations` : Number of training epochs
- `l1_reg` : L1 regularisation parameter
- `l2_reg` : L2 regularisation parameter
- `learning_rate` : Learning rate in the Adam optimiser
- `hidden_layer_sizes` : Number of neurons in each hidden layer
- `batch_size` : Size of the mini-batches used in the optimisation
- `n_basis` : Number of values to use in $R_s$ and $\theta_s$ in the ACSFs
- `r_min` : First value of $R_s$
- `r_cut` : Cut-off radius in the ACSFs
- `tau` : Parameter used to calculate η and ζ in the ACSFs

The hyper-parameter `tau` (τ) is used to calculate η and ζ as follows:

$$\eta = \frac{4 \cdot \log(\tau) \cdot (N_{\text{basis}} - 1)^2}{(R_c - r_{min})^2} \tag{S1}$$

$$\zeta = -\frac{\log(\tau)}{2 \cdot \log\left(\cos\left(\frac{\pi}{4N_{\text{basis}} - 4}\right)\right)} \tag{S2}$$

*Table 1 - Hyper-parameters used when training the ANNs on the different mixed data sets.*

| Parameter | Training set 1 | Training set 2 | Training set 3 | Training set 4 | Training set 5 | Training set 6 |
|---|---|---|---|---|---|---|
| `iterations` | 639 | 1338 | 965 | 1181 | 1424 | 900 |
| `l1_reg` | 1.5e-4 | 6.4e-7 | 1.0e-6 | 2.5e-4 | 1.5e-6 | 1.9e-4 |
| `l2_reg` | 3.5e-7 | 2.2e-5 | 4.2e-7 | 4.1e-5 | 8.7e-5 | 2.2e-8 |
| `learning_rate` | 2.3e-3 | 1.2e-3 | 4.3e-4 | 7.1e-4 | 7.0e-4 | 1.5e-3 |
| `hidden_layer_sizes` | (272,179) | (393,154) | (280,326) | (94,174) | (235,144) | (62,142) |
| `batch_size` | 23 | 24 | 22 | 26 | 43 | 23 |
| `n_basis` | 15 | 19 | 13 | 13 | 12 | 16 |
| `r_min` | 0.8 | 0.8 | 0.8 | 0.8 | 0.8 | 0.8 |
| `r_cut` | 4.3 | 3.4 | 3.6 | 3.4 | 3.7 | 3.1 |
| `tau` | 1.7 | 1.4 | 1.7 | 1.9 | 2.2 | 1.8 |

## S2. Reproducing our results

We have created a repository with instructions on how to reproduce our results. The repository can be found at

https://github.com/SilviaAmAm/squalane_paper_si